# Retinal arterial blood flow measured by real-time Doppler holography at 33,000 frames per second

Yann Fischer, Zacharie Auray, Olivier Martinache, Marius Dubosc, Noé Topéza, Chloé Magnier, Maxime Boy-Arnould, Michael Atlan

Langevin Institute - CNRS. École Supérieure de Physique et de Chimie Industrielles (ESPCI) Paris. Paris Sciences & Lettres (PSL) University. 1, Rue Jussieu. 75005 Paris, France

*Abstract*—This study presents a novel quantitative estimation method for total retinal arterial blood flow utilizing real-time Doppler holography at an unprecedented frame rate of 33,000 frames per second. This technique, leveraging high-speed digital holography, enables non-invasive angiographic imaging of the retina, providing detailed blood flow contrasts essential for assessing retinal health. The proposed quantitative analysis method consists of segmenting primary in-plane retinal arteries and calculating local blood velocity using Doppler frequency broadening. The analysis integrates a forward scattering model to achieve blood flow estimation. Our findings highlight the potential of Doppler holography as a powerful tool for diagnosing and monitoring the treatment of retinal vascular conditions, complementary to existing imaging methods.

## I. Introduction

The retina, composed of ten neuronal layers including photoreceptor, neuronal, and glial cells, is a thin structure (0.1 to 0.3 mm) located within the posterior pole of the eye. It receives its blood supply primarily from the central retinal artery, which divides into branches that maintain distinct paths across the retina, forming four vascular networks. Retinal blood flow is tightly autoregulated to ensure stability despite fluctuations in cardiac output, adjusting vessel diameter to meet the metabolic needs of the tissue. While the retina receives 20-30% of the total ocular blood flow, the majority is directed to the choroid, which supports the outer retina, including the photoreceptors. The choroid, with its high vascular density, has one of the highest blood flows per unit weight of any tissue, essential for temperature regulation, nutrient delivery, and waste removal. Unlike the retina, which shows efficient autoregulation, choroidal circulation lacks autoregulatory mechanisms [1].

The retinal vasculature plays a critical role in many ocular diseases that lead to vision loss. While fluorescein angiography remains the gold standard for evaluating retinal vessels, it carries risks of adverse effects and fails to image all retinal layers comprehensively. Optical coherence tomography angiography (OCT-A) offers en face imaging of retinal vasculature but the ability to capture flow dynamics is focused on flow in retinal capillaries and uses a surrogate marker for blood flow speeds [2-4]. Laser Doppler flowmetry (LDF) provides blood flow measurements in specific retinal regions but suffers from poor spatial resolution, making it difficult to distinguish between retinal layers or specific vascular structures [5]. Similarly, Laser Speckle Contrast Imaging (LSCI) offers only relative blood flow measurements and is highly sensitive to motion artifacts, which can distort results [6]. Ultrasound Doppler methods, though traditionally used to assess retrobulbar blood flow, are limited in clinical application due to insufficient spatial resolution and regulatory intensity limits for retinal imaging [7].

Despite recent advancements in imaging techniques like indocyanine green angiography, fluorescein angiography, and OCT angiography, these methods still struggle to capture blood flow dynamics throughout the cardiac cycle. To better manage ocular conditions, innovative approaches are needed. Doppler holography [8-12], a promising non-invasive technique, addresses this by offering high temporal resolution imaging of retinal and choroidal blood flow. Designed for large-scale patient screenings, it uses near-infrared, high-speed digital holographic imaging to reliably estimate local blood flow in retinal vessels. The integration of a co-designed optical interferometry system and high-performance computing in Doppler holography aims to identify new biomarkers for improved non-invasive pathology classification and therapeutic monitoring.

## II. Proposed analysis method for quantitative assessment of retinal blood volume rate using real-time Doppler holography signals

Doppler holography uses laser light for interferometric measurements of the retina, combining holography, which captures the phase and amplitude of reflected light, with the Doppler effect, detecting frequency shifts from moving blood cells. A laser illuminates the retina, and a high-speed camera records the interference pattern, creating detailed images of endoluminal blood flow contrasts. One notable feature of Doppler holography is its lack of depth sectioning, which allows for choroid visualization across the field of view but prevents straightforward retinal blood flow estimation through Doppler flowmetry analysis methods [5].

The proposed analysis method for quantitative estimation of total retinal blood flow using Doppler holography can be achieved through a straightforward and robust process that leverages the retina's inherent optical and topological properties. The method consists of two key steps:

1. The primary in-plane retinal arteries are segmented, and the differential Doppler frequency broadening is measured relative to the surrounding tissue. When the local neighborhood signal does not accurately reflect the local retinal vessel background, this can lead to either underestimation or overestimation of the local differential Doppler broadening. These discrepancies, including the sign of the frequency difference, are preserved, and statistically, variability in these measurements tends to cancel out across the field of view.

2. A forward scattering model is applied (Fig. 1), incorporating a secondary diffused light source from the sclera that passes through the choroid and deeper retinal layers. This model enables the calculation of blood flow velocity in the retinal arteries, proportional to the local differential Doppler broadening. The only parameters required for velocity estimation are the optical wavelength and the eye's numerical aperture.

Finally, artery sections are selected from their distance to the central retinal artery and the local average velocity in each section is multiplied by the cross-sectional area, calculated from an estimated radius, to obtain blood volume rate estimations in each retinal artery sections considering they present a Poiseuille velocity profile.

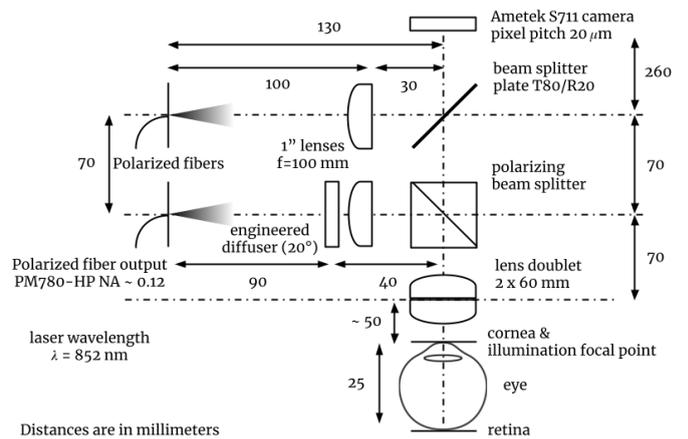

Fig. 2. Illustration of the optical setup. An inline Mach-Zehnder interferometer operating in the near-infrared range combines light backscattered from the patient's eye fundus with a reference beam. The resulting optical interference patterns are captured in real time by a streaming camera.

### III. DEVICE SETUP, EXPERIMENTAL PROTOCOL AND SIGNAL ACQUISITION

The experimental setup utilizes a Mach-Zehnder inline interferometer (Fig. 2). Near-infrared radiation from a diode laser (Thorlabs FPV852P, wavelength: $\lambda = 852$ nm) is split into linearly-polarized reference (10%) and illumination (90%) arms, using polarization-maintaining fibers (Thorlabs PM780-HP, numerical aperture: NA ~ 0.12). The illumination beam is diffused through an engineered diffuser (Thorlabs ED1-C20-MD, 1" diameter, 20° circle tophat) and focused by an eyepiece composed of two biconvex lenses with a combined effective focal length of ~33 mm. A real-time camera (Ametek Phantom S711 with Euresys Coaxlink QSFP+ frame grabbers, frame rate: 33 kHz, pixel pitch: 20 µm, frame size: 384×384 pixels) captures the images. The retina of a volunteer is illuminated by the diffuse laser beam focused through the eyepiece as shown in Fig. 1 and Fig. 2. The cross-polarized backscattered light interacts with the reference beam, creating interferogram patterns that are captured by the high-speed camera [11].

Informed consent was obtained from the subject, and the clinical investigation received authorization from the relevant ethics review boards, including the Personal Protection Committees (South-West and Overseas III) and the National Agency for the Safety of Medicines and Health Products (ANSM). The study was registered under the reference IRDCB 2023-A00316-39-A. The volunteer's positioning was carefully monitored through real-time computation and visualization of inline digital holograms of the eye fundus, generated from a 16-bit, 384×384-pixel interferogram stream recorded at 33,000 frames per second using the Phantom S711 camera. This was achieved using Fresnel transformation and principal component analysis on stacks of 16 consecutive holograms with the Holovibes digital hologram streaming software [9]. A total of 131,072 raw interferogram frames, amounting to 18 GB of data, were acquired during real-time image rendering and later used for offline image rendering and analysis. The imaging prominently captured the optic nerve head and surrounding areas, clearly displaying the primary branches of the central retinal artery, which supplies the retina, and the central retinal vein, which drains it (Fig. 3).

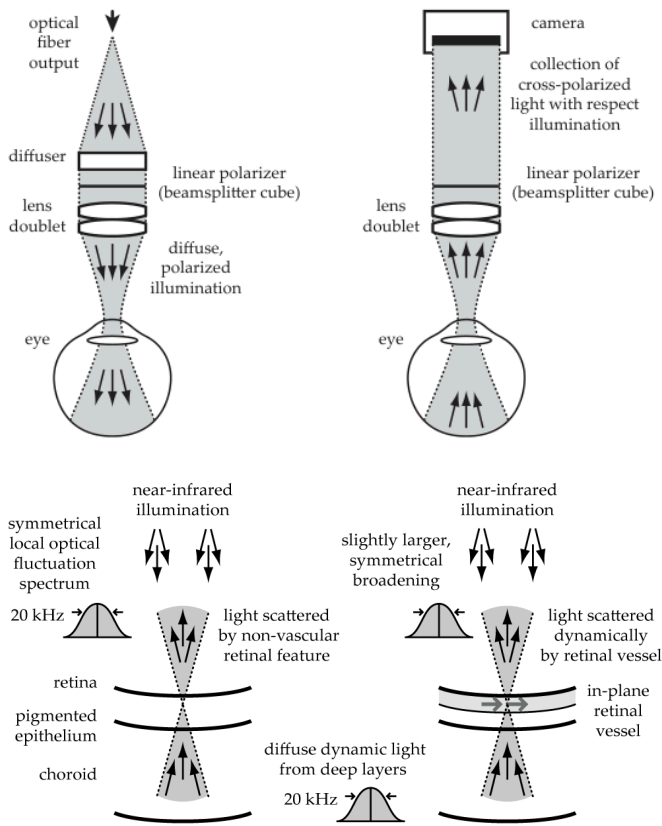

Fig. 1. Illumination (top left) and detection (top right) conditions using linearly polarized near-infrared coherent light. A forward scattering model accounts for a diffuse, Doppler-broadened secondary light source within the sclera and deep choroid to estimate blood flow in primary in-plane retinal vessels. The presence of these vessels induces a subtle, symmetrical Doppler broadening of the secondary light source, as depicted in the bottom sketches. This measurable broadening is utilized to estimate local endoluminal root-mean-square (RMS) blood velocity.

## IV. Image rendering and signal analysis

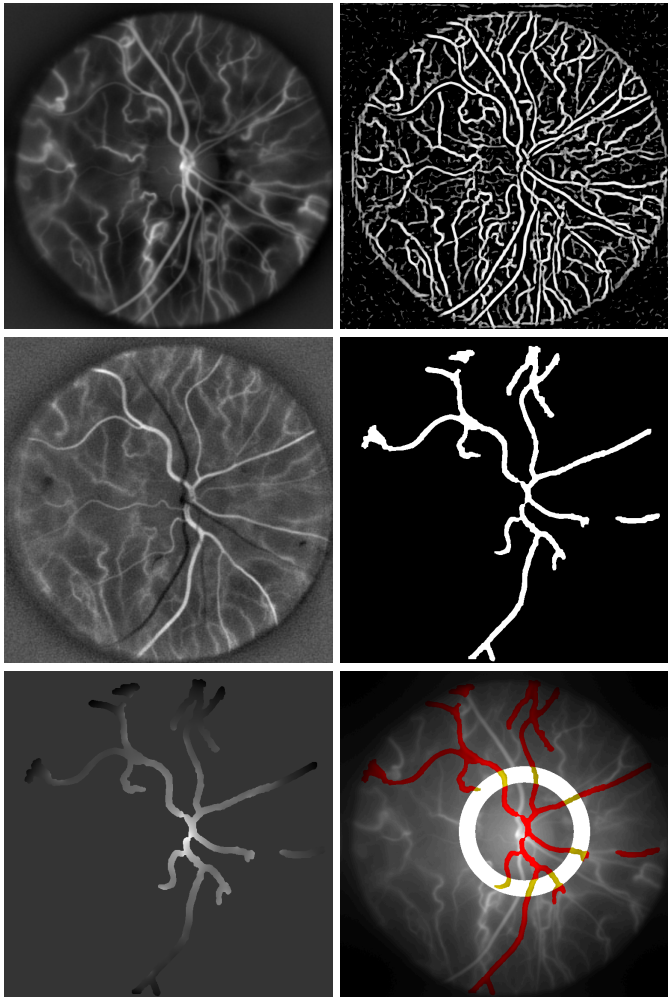

Fig. 3. From left to right, top to bottom: Power Doppler image averaged over four cardiac cycles, highlighting blood flow contrasts in the retina and choroid. Frangi segmentation [13]. Temporal correlation map with spatially averaged signals revealing retinal arteries. Segmented retinal arteries. Local background estimate from the neighborhood of the segmented arteries. Composite power Doppler image showing segmented retinal arteries in red.

Offline image rendering of the raw interferograms was accomplished using Fresnel transformation, singular value decomposition filtering, and short-time Fourier transform applied within 512-frame windows to generate power Doppler images [10].

The retinal arteries were isolated from the power Doppler images through a series of segmentation steps. First, a flat-field correction was applied to correct for non-uniform illumination and normalize the intensity across the field of view. Next, Frangi vessel segmentation [13] — a well-established technique in ophthalmology — was used on the locally contrasted power Doppler images to map the vessels of the eye fundus. Temporal correlation between the vessel mask and the zero-mean spatially averaged signal was then computed to enhance the visibility of the arteries. A manually adjusted threshold was applied to create a preliminary artery mask. Finally, the segmented retinal arteries were refined and connected using filtering methods focusing on the most connected areas. Choroidal vessels distant from the optic nerve were removed during this step as long as they were not too close to retinal vessels.

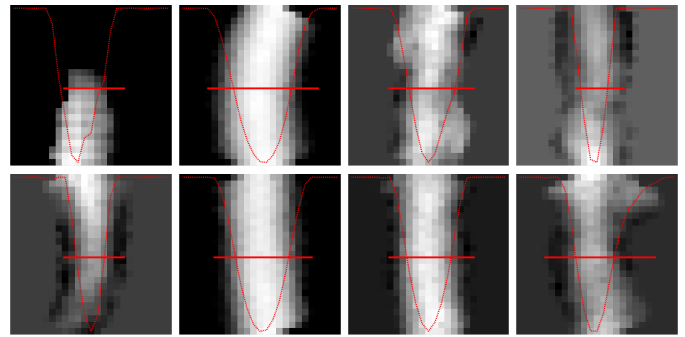

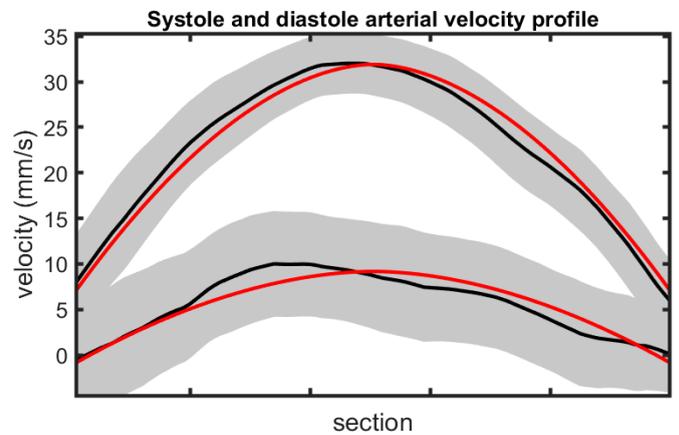

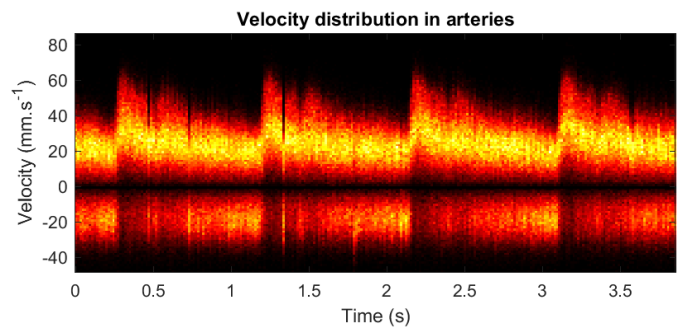

Fig. 4. Top : Local artery sections used to compute the average RMS velocity profile (displayed in red). Middle: wall-to-wall arterial velocity profile averaged across selected artery sections at diastole (lower profile) and systole (upper profile); the gray area represents one standard deviation of the estimated velocity signal along the flow direction. Bottom : The estimated velocity distribution of arterial blood flow shows a range from positive to negative values, revealing a statistical discrepancy: the measured local Doppler spectrum within a retinal artery can be narrower than that of its local background estimate from the neighborhood of the segmented arteries.

Blood velocity in segmented retinal arteries was estimated locally by measuring differential Doppler frequency broadening relative to surrounding areas. This locally increased broadening in retinal arteries, denoted as $\Delta f$, was estimated by calculating the square root of the absolute difference between the local normalized second-order moments of the Doppler spectrum with the background estimated from its neighborhood (Fig. 3) within the 6 kHz to 16.5 kHz time frequency band, across the field of view. The result was then multiplied by +1 or -1, depending on whether the sign of this difference was positive or negative, to account for instances where the signal from the local neighborhood does not accurately estimate Doppler broadening in the arterial background. This approach yielded estimated velocity distributions with a negative component (Fig. 4). The hypothesis here suggests that the local surroundings experience the same optical path and Doppler broadening as the light under the artery, isolating global Doppler broadening from vessel motion-related broadening thanks to segmentation. However, this assumption is not accurate locally in practice, leading to over- or underestimation. To correct these discrepancies, a sign function is applied to the root-mean-square (RMS) difference, which helps balance out the local errors on average, over the field of view. We assume that the discrepancy introduced by local artery neighborhood measurements is statistically negligible. A forward scattering physical model (Fig. 1), based on a diffused secondary light source in deeper retinal layers, was employed to derive the local RMS blood flow velocity $v$ in in-plane retinal arteries from the estimated local increase in Doppler broadening in retinal arteries $\Delta f$. This local velocity was calculated as the product of the optical wavelength ($\lambda$ = 852 nm) and the local differential Doppler broadening ($\Delta f$), divided by the numerical aperture (NA) of the eye, estimated to be 0.124,

$$v = \lambda \, \Delta f \, / \, \text{NA}. \quad (1)$$

Artery sections were selected from branches coming from the central retinal artery intersecting with a circle of adjustable radius and width (Fig. 3, bottom right) then they were rotated to find the maximum value of the vertical summation and finally velocity profiles averaged along the section width were computed (Fig. 4, top). The in-plane length per pixel was determined using a known estimate of the papilla diameter. Cross section areas were then calculated from the arteries radius estimated from the velocity profile in each branch. These velocity profiles were combined in Fig. 4 (center) to compare the average wall-to-wall velocity profile across arteries during diastole and systole, and fitted to a Poiseuille profile (in red). Blood viscosity increases as shear rate decreases, causing resistance to flow and raising pressure. Higher viscosity results in reduced flow at a given pressure. Blood viscosity is affected by factors like red blood cell deformability, hematocrit, and plasma proteins. In vivo and in vitro studies indicate that blood velocity profiles in retinal arteries, especially those under 100 micrometers in diameter, may be blunter than the ideal parabolic shape and can vary throughout the cardiac cycle [14, 15].

The local absolute blood volume rate was determined by multiplying the local velocity (Eq. 1) by the cross-sectional area of the segmented principal retinal arteries, at arbitrary locations within a given radius from the centroid of the segmented vessels. The total arterial blood volume rate and arterial resistivity index were estimated by summing the measurements across the principal retinal arteries (Fig. 5). The average total retinal blood volume rate observed in our control subject is 35 μL/min (Fig. 5).

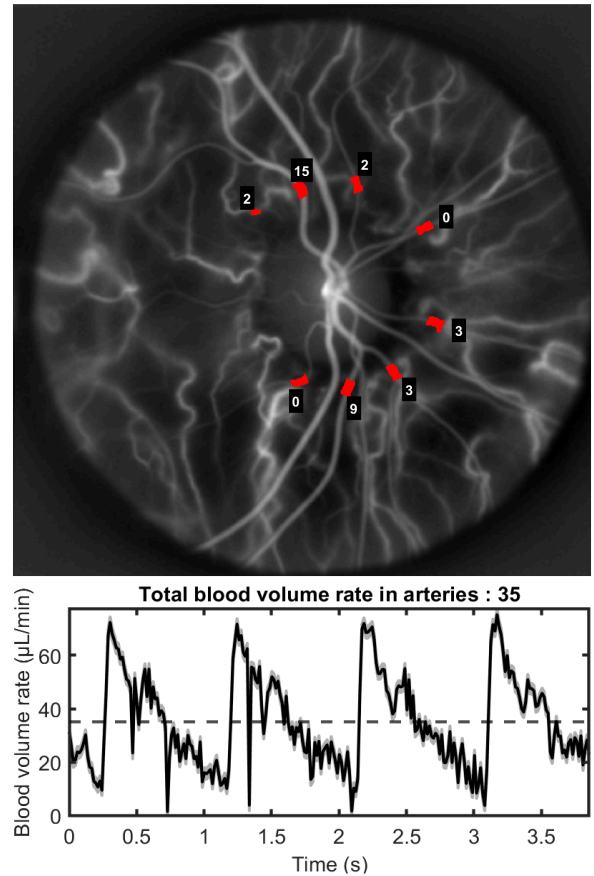

Fig. 5. Arterial blood volume rate estimated in a volunteer from 3.8 s of interferograms recorded at 33,000 frames per second.

V. SOFTWARE AND DATA

The software developed for the estimation of total retinal arterial blood flow using real-time Doppler holography is open-source and available on GitHub: It includes real-time, concurrent image rendering and raw image saving, offline image rendering and Doppler lineshape measurement, and offline segmentation and retinal blood flow estimation. Image rendering and analysis were performed using the following software versions: [Holovibes](.) Release 13.2.3, [HoloDoppler](.) Release 1.2, and [Pulsewave](.) Release 1.4. The raw dataset used for the reported results is available upon request.

VI. CONCLUSION

The proposed deterministic blood flow analysis routine, utilizing real-time Doppler holography at 33,000 frames per second, may still underestimate the total retinal blood volume rate because the current recording frame rate is insufficient to fully capture the local Doppler broadening in principal retinal vessels. This causes some signal clipping above the Nyquist frequency when blood velocity increases during the systolic peak [8]. Increasing the recording frame rates could potentially resolve this issue. Yet the assessed retinal blood volume rate presented in this study of 35 μL/min is of the same order as the reported average total arterial and venous volumetric flow rates in the literature are 33 ± 9.6 μL/min and 34 ± 6.3 μL/min, respectively, as measured in 12 eyes using bidirectional laser Doppler velocimetry [16]. Additionally, Doppler FD-OCT measurements in eight out of ten subjects reported a mean (SD) total retinal blood flow of 45.6 (3.8) μL/min, with a range of 40.8 to 52.9 μL/min [17].

The proposed approach for quantitative estimation of retinal blood flow using real-time Doppler holography shows promise for detecting and monitoring retinal diseases. Its ability to provide absolute measurements of retinal blood supply could significantly enhance treatment monitoring for conditions like diabetic retinopathy and glaucoma. Furthermore, this technique could be crucial for early detection and monitoring of cardiovascular conditions, including atherosclerosis and hypertension. Enhancing automatic segmentation reliability will be key to fully realizing its clinical applications. This technology has the potential to transform ocular and cardiovascular health monitoring, seamlessly integrating into routine clinical practice and improving patient care.

ACKNOWLEDGMENT

Funding : ANR LIDARO ANR-22-CE19-0033-01.